\documentclass[epj]{svjour}

\usepackage{epsfig}

\begin{document}

\title{Dynamics and Scaling of Noise--Induced Domain Growth}

\author{M. Iba\~nes\inst{1} \and J. Garc\'{\i}a-Ojalvo \inst{2}
\and R. Toral \inst{3} \and J.M. Sancho\inst{1}}

\institute{
Departament d'Estructura i Constituents de la Mat\`eria,
Universitat de Barcelona, Diagonal 647, E--08028 Barcelona, Spain
\and
Departament de F\'{\i}sica i Enginyeria Nuclear, Universitat
Polit\`ecnica de Catalunya, Colom 11, E--08222 Terrassa, Spain
\and
Instituto Mediterr\'aneo de Estudios Avanzados, IMEDEA (CSIC-UIB), Campus
UIB, E-07071 Palma de Mallorca, Spain
}
\date{\today }

\abstract{
The domain growth processes originating from noise-induced
nonequilibrium phase
transitions are analyzed, both for non-conserved and conserved
dynamics. The existence of a dynamical scaling regime is established
in the two cases, and the corresponding growth laws are determined.
The resulting universal dynamical scaling scenarios are 
those of Allen-Cahn and Lifshitz-Slyozov, respectively.
Additionally, the effect of noise sources on the behaviour
of the pair correlation function at short distances is studied.
}

\maketitle

\section{Introduction}

Growth processes for systems in their evolution towards a final state of
thermodynamic
equilibrium have been intensively studied during the last decades \cite{gsms}.
A particularly interesting situation concerns the case in which equilibrium
corresponds to either of two equivalent phases of a scalar field
$\phi(\vec x,t)$. Since in equilibrium the relevant
free energy $F$ is minimized, ``equivalent" stands in this context for
phases that have the same (minimum) value of that free energy. A typical
case corresponds to the evolution following a sudden quench from
a very high-temperature, homogeneous phase, to a final state below a critical
temperature for an order--disorder transition.
In this process, one can distinguish three regimes:

i) Right after
the quench, the system finds itself out of equilibrium and small
domains start to appear, corresponding to islands of each one of the
two phases. The initial instability mechanism varies with the system.

ii) At late times, motion of the domain walls separating
equivalent phases
leads to a situation of domain growth, also known as {\em coarsening}.
This regime occurs when the typical domain size $R(t)$ is much
larger than the width of the interface between domains and much smaller
than the system size.
The main general mechanism for coarsening  is the
reduction of the excess of interfacial energy, which is accomplished by
reducing the interface curvature. The exact details of the domain growth
dynamics depend on the spatial dimension $d$ and on the existence (or
absence) of conservation laws. For dimension $d>1$ and for a system whose
order parameter is not conserved (i.e., which evolves towards a final
one-phase, symmetry--breaking, equilibrium ordered state), the 
domain size $R(t)$ grows with time $t$ as ${\displaystyle R(t)\sim t^{1/2}}$
(Allen-Cahn law).
On the other hand, for a system with a conserved order parameter (i.e.,
in which the final state has  two coexisting phases), it
has been found
that ${\displaystyle R(t)\sim t^{1/3}}$ (Lifshitz-Slyozov law).
Moreover, in both cases the relaxation of these systems
towards equilibrium has been found to be self-similar. This claim basically
states that, after the initial transient regime, the only relevant length scale
is the domain size $R(t)$, so that two images of the system taken at two
different times, $t_1$ and $t_2$, can be made to match (in a statistical
sense) by rescaling their lengths by $R(t_1)$ and $R(t_2)$, respectively.
The validity of this {\sl dynamical scaling law} \cite{bs74,lmk82} has a
precise statement in terms of the correlation function, and has been
extensively verified in many experiments, as well as in computer
simulations of discrete and continuous models \cite{gsms}.

iii) Finally, when the domain size $R(t)$ is of the order of the system size,
the system asymptotically reaches
an equilibrium steady state, in which either only one of the two
phases occupies the whole system (non-conserved order parameter case)
or two large domains separated by a single boundary coexist
(conserved order parameter case). In this case, the only relevant
length scale of the problem is the equilibrium static correlation length.

Quite generally, the growth processes mentioned above have been described
theoretically by the following dynamical equation:
\begin{equation}
\label{modelA}
\frac{\partial\phi(\vec x,t)}{\partial t}=-\Gamma\frac{\delta {\cal F}}
{\delta \phi(\vec x,t)}+\eta(\vec x,t)\,,
\end{equation}
where ${\cal F}[\phi]$ is a suitable nonequilibrium free-energy
functional, $\Gamma$ is a kinetic coefficient and $\eta(\vec x,t)$ is a
Gaussian white noise representing internal fluctuations, taken to have
zero mean and correlation:
\begin{equation}
\langle \eta(\vec x,t)\,\eta(\vec x',t')\rangle=2\,\varepsilon\,
\delta(\vec x-\vec x')\,\delta(t-t')\,.
\label{aditivoA}
\end{equation}
When $\epsilon=\Gamma k_B\,T$ (what is known as the fluctuation-dissipation
relation),
the system reaches an equilibrium state governed by the well-known
Boltzmann--Gibbs
distribution $\exp[-{\cal F}/k_B\,T]$. In the absence of fluctuations,
the free--energy ${\cal F}$ is a Lyapunov potential for the dynamical
evolution, and the resulting dynamics is classified as potential \cite{poten}.
It is worth stressing that the growth laws and dynamical scaling
introduced before hold independently of the strength of the fluctuation
terms, i.e. of the value of $\epsilon$. This is true as far as those
fluctuation terms are not extremely large (such that they inhibit the
formation of order), or near a critical point (where fluctuations are
largely amplified).

Most of the work in the past has dealt with the simpler case when the
corresponding deterministic dynamics is potential.  
Only in last years, attention has been focused on more general
dynamics where the final state is not of thermodynamic equilibrium.
Most studies have been concerned with the case in which the deterministic
dynamics does not imply the minimization of a free energy, the so--called
non-potential dynamics\cite{nonpot}. New and interesting effects can
appear in this case. For instance, it has been found that several stable
phases can coexist in the system in a non-coarsening situation
\cite{nonpot2}. 

A new type of non-potential systems presenting order--disorder phase
transitions has been introduced recently. The main feature of those systems
is that the initial instability leading to domain growth and coarsening is
induced by fluctuation terms of external origin. This is a rather
counterintuitive effect of noise, in that an increase of the noise intensity
can lead to a spatially ordered pattern. This phenomenon is nowadays well
established and has lead to a host of new effects in which noise has a
spatial ordering role (see \cite{nises} for a
recent review). These effects include
enlargement of the coexistence region in standard models of phase transitions
\cite{pla}, pure noise--induced phase transitions \cite{vpt},
noise--driven structures in pattern--formation processes \cite{prl}, and
noise-sustained pulse propagation \cite{epl}, among others.
In all these cases, a nonequilibrium ordered pattern can emerge as
a consequence of the fluctuating terms, and disappears when the noise
source is switched off. The final state cannot be made to correspond to
any known free-energy, and the equivalence of the different phases
can be established, whenever possible, by symmetry arguments. These
systems can be described by the following Langevin equation 
\begin{equation}
\label{laneq}
\frac{\partial\phi(\vec x,t)}{\partial t}=-\Gamma\left[\frac{\delta
{\cal F}}{\delta \phi(\vec x,t)} + g(\phi)\xi(\vec x,t)\right]+
\eta(\vec x,t)\,,
\end{equation}
where $\xi(\vec x,t)$ represents external fluctuation sources
(multiplicative noise, if $g$ depends on $\phi$),
to be characterized statistically (see later).

So far, studies of noise-induced spatial order have been concerned with
the characterization of the final, nonequilibrium steady state
(which would correspond to regime  (iii) above).
In particular,
much effort has been devoted to determine the phase diagram and the critical
properties of these systems, including the possible existence of new
universality classes for the steady state. A great deal of our knowledge
of the detailed behaviour of these systems comes from numerical simulations
of the corresponding stochastic partial differential equations. On the
theoretical side, studies
have used mean--field theories, dynamic renormalization group,
and linear stability analyses \cite{nises}.
Renormalization-group arguments, for instance, show that multiplicative
noise is less relevant than additive noise, when both are present, to the
universal behaviour,
and thus it does not change the universality of the resulting
critical points \cite{nises}.
In this paper, on the other
hand, we are concerned with the noise-induced
{\em dynamical} evolution of both
conserved and non-conserved systems towards a nonequilibrium steady state
(which would correspond to the scaling regime (ii) defined
above).
We will be interested in studying
the growth laws and whether dynamical scaling holds in these cases. 

This paper is organized in the following way. In section II we study numerically
the scaling properties and the growth of domains for the two--dimensional
{\em non-conserved} Ginzburg--Landau model in presence of both 
external and internal fluctuations. In section III we characterize numerically
the phase separation dynamics induced by external noise in the
two-dimensional {\em conserved} Ginzburg--Landau model.
Finally, section III is devoted to the study of the pair
correlation function at short distances in the presence of noise sources.

\section{Non-conserved model}

As stated in the introduction, when a system in a high--temperature disordered
equilibrium state is quenched below its order--disorder transition temperature,
for which equivalent ordered phases can coexist,  domains of the new
equilibrium phases appear and grow.  For those systems whose order parameter is
not conserved, one of the domains grows until it fills the whole system,
assumed finite. The
domain boundaries move with a translational velocity that has been found,
for spatial dimension $d\ge 2$, to be basically proportional
to the mean curvature of the boundaries, and independent of the free energy of
the interface. This mechanism leads to the well-known Allen--Cahn law for the
growth of the average linear size $R(t)$ of phase domains:
$R(t)\propto t^{1/2}$. The same law, with a different physical
mechanism, has been shown to hold in other models as well \cite{rafa00}.

A particular model for which the above results have been established is
the non-conserved Ginzburg--Landau model (known as model A in the
literature of
critical phenomena\cite{hh}). It is defined by Eq.~(\ref{modelA}) with
a constant kinetic coefficient $\Gamma=1$, and with the following free energy:
\begin{equation}
\label{fef}
{\cal F}=\int d\vec x \left[-\frac{a}{2}\phi^2+\frac{1}{4}\phi^4+
\frac{D}{2}|\vec\nabla\phi |^2\right]\,.
\end{equation}
For fixed values of the diffusion constant $D$ and the internal noise
strength $\epsilon$, there is a critical value $a_c(D,\epsilon)$ such
that for $a<a_c$ equilibrium corresponds to a single homogeneous phase
(the {\sl disordered} phase), whereas for $a>a_c$ two equivalent equilibrium
phases (with the same minimum value of the free energy) coexist. Due to the
symmetry of $\cal F$, the {\sl ordered} phases are related by a global
change of sign of the field $\phi$. The critical value $a_c$ vanishes for
zero noise intensity $a_c(D,\epsilon=0)=0$. 

A new ordering situation can arise if we let fluctuate the control parameter
$a\to a+\xi(\vec x,t)$. Even when the mean value of $a$ is still
below the critical level $\langle a \rangle < a_c$, it might be
possible, by choosing a sufficiently large intensity of the external
fluctuations, to have a bistable stationary (but no longer
equilibrium) distribution for the field. The two maxima of the stationary
distribution are again symmetric under a global change of sign. 

\begin{figure}[htb]
\begin{center}
\epsfig{file=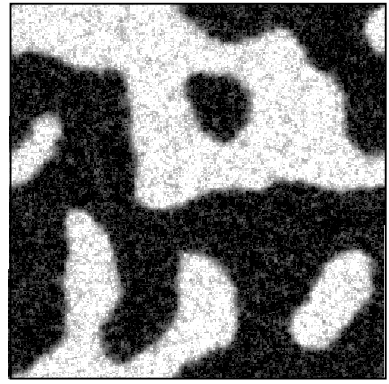, width=2.7in}
\end{center}
\caption{Spatial pattern for the stochastic model A with
$\sigma^2=0.4$, $\epsilon= 10^{-4}$,
$a=-0.2$, and $D=1$ at t=200. The square lattice has
$256\times 256$ cells of mesh size $\Delta x=1$. The black and white domains 
correspond to symmetric phases.} 
\label{fig:patt}
\end{figure}

In the presence of both internal and external fluctuations, the
non-conserved Ginzburg--Landau model is obtained by substitution of the
free energy functional (\ref{fef}) in the Langevin equation (\ref{laneq}).
One finds
\begin{equation}
\frac{\partial \phi(\vec x,t)}{\partial t} =a\phi-
\phi^3 +\,D\,
\nabla^2\phi+ \phi\,\xi(\vec x,t) + \eta(\vec x,t)\,,
\label{LangevinA}
\end{equation}
where both additive and multiplicative noises are Gaussian, with zero mean.
The additive-noise correlation is defined by Eq.~(\ref{aditivoA}).
In general, the external multiplicative noise $\xi(\vec x,t)$ may have
specific temporal and spatial correlation functions. For the moment,
we will assume that:
\begin{equation}
\langle \xi(\vec{x},t)\,\xi(\vec{x}',t')\rangle=2\,\sigma^2\,
c(\vec x-\vec x')\,\delta(t-t')\,,
\label{multiplicativoA}
\end{equation}
i.e. a delta--correlated function in time and a general correlation function $c(\vec x)$ in space.
In this case, for given $\epsilon$ and $D$, the critical value of $a$ depends
on $\sigma$ in such a way that its value is {\em lowered} with respect to the
case with no external fluctuations, $\sigma=0$. In the limit of small
$\epsilon$, a linear stability analysis \cite{becker94,jordi98}
and a mean--field type theory \cite{chris,Ibanes99}
give the approximate result $a_c=-\sigma^2c(0)$. Since
ordered states appear for $a>a_c$, by choosing parameter values
satisfying  $-\sigma^2c(0)<a<0$ we have a situation in which nonequilibrium
steady phases {\sl induced by external noise} appear and grow (see
Fig. \ref{fig:patt}).
If external fluctuations are switched off the ordered state disappears,
since the (negative) value of $a$ is now below the critical value. In
this section, we want to elucidate if the growth of these domains still
follows the Allen--Cahn law and if the system still exhibits a scaling regime.  

Most of the results of this section have been obtained by a
numerical integration of the stochastic model defined by Eq.~(\ref{LangevinA}).
In order to perform the numerical analysis, we redefine the model by
considering a regular d--dimensional lattice with $N^d$ points and
lattice spacing $\Delta x$:
\begin{equation}
\frac{d \phi_i}{ d t} = a\,\phi_i - \phi_i^3+\!D \sum_{j}
\,\widetilde D_{ij}\,\phi_j + \eta_i(t)\, + \phi_i\,\xi_i(t)\,,
\label{eq:spdedis}
\end{equation}
where $\phi_i(t) \equiv \phi(\vec x_i,t)$ and only one index is used to label 
the cells, independently of the dimension of the lattice.
$\widetilde D_{ij}$ accounts for the discrete Laplacian operator
\begin{equation}
\label{eq:laplaop}
\nabla^2 \,\to \,\sum_j\,\widetilde D_{ij}=\frac{1}{(\Delta x)^2}\sum_j\left(
\delta_{nn(i),j}-\,2d\,\delta_{i,j}\right)\,,
\end{equation}
where $nn(i)$ represents the set of all the sites that are
nearest neighbors of site $i$.
The discrete noises $\eta_i(t)$ and $\xi_i(t)$
are still Gaussian with zero mean and correlations
\begin{equation}
\langle \eta_i(t)\,\eta_j(t')\rangle=2\,\varepsilon\,
\frac{\delta_{i,j}}{(\Delta x)^d} \,\delta(t-t')
\label{eq:wnc2dis}
\end{equation}
and
\begin{equation}
\label{eq:wmn2dis}
\langle \xi_i(t)\,\xi_j(t')\rangle=2\,\sigma^2\,
\frac{\delta_{i,j}}{(\Delta x)^d} \,\delta(t-t')\,.
\label{eq:ncmcold}
\end{equation}
This last expression
corresponds to an external noise with a white spatial correlation
in the lattice, i.e. $c(0)=1/\Delta x^d$. Since we will choose $\Delta x=
1$ in all the results shown in the paper, $c(0)=1$ in what follows.

After the discretization procedure described above,
the resulting set of coupled stochastic differential equations for the
variables $\phi_1,\dots,\phi_N$ has been integrated numerically. 
We have studied two cases of noise--induced non-conserved dynamics,
corresponding to two different multiplicative noise intensities, 
$\sigma^2=0.4$ and  $\sigma^2=0.6$. In both cases, the additive noise
intensity is $\epsilon=10^{-4}$, and $a=-0.2$. Since
these values satisfy $-\sigma^2 < a<0$, the linear
stability analysis indicates that the homogeneous state is unstable for
these multiplicative noise intensities \cite{becker94,jordi98}. We have
compared these two cases with the purely deterministic
model $\varepsilon=\sigma=0$ for $a=1$, and with the stochastic model with
only additive noise ($\sigma=0$) for $a=2$ and $\epsilon=0.7$. In all
cases we have chosen the coupling coefficient to be $D=1$.

Our numerical analysis has been performed in a 
two--dimensional square lattice of $256\times 256$ 
cells with periodic boundary conditions and mesh size $\Delta x=1$,
except for the stochastic model with only additive noise, which has
been considered in a lattice of $128\times 128$ cells in order to compare
with the results of Ref. \cite{Oguz90}. We have used a first--order Euler
algorithm with a time step $dt=10^{-2}$ in the deterministic case, and
$dt=5\cdot 10^{-3}$ in presence of noise sources. The noises have been
generated using a numerical inversion method \cite{Tor94}. As initial
condition, we have chosen in all cases a random uniform distribution of
the field in the interval $(-1,1)$ in order to simulate a high--temperature 
one--phase state. Our results are averaged over 15 samples in the
deterministic case, 40 samples in the purely additive-noise case, and
30 samples in the two multiplicative-noise cases.

In order to study the growth process and to examine the existence of
dynamical scaling, we define the correlation and structure functions.
These two functions give information about the spatial
structure of the system at a certain time. By knowing how these functions
evolve, one can obtain information about the growth of domains, and check if
there is a scaling
regime. In our discrete space, we define the pair correlation function as:
\begin{equation}
G(\vec{r}_j,t)=\left\langle \frac{1}{N^d}\sum_{i}\phi(\vec{r}_j+
\vec{x}_i,t)\,\phi(\vec{x}_i,t)\right\rangle
\label{eq:corrA}
\end{equation}
where the brackets denote an average over different initial conditions 
and realizations of the noises.

The structure function is the (discrete) Fourier transform of the pair
correlation function: 
\begin{equation}
S(\vec{k}_{\mu},t)=(\Delta x)^{d}\sum_{j}e^{-i\vec{r}_j\cdot \vec{k}_{\mu}}
G(\vec{r}_j,t)\,,
\label{s:tfg}
\end{equation}
where $\mu$ labels the $N^d$ points in Fourier space.

In practice, the structure function is computed in an easier way
using the equivalent definition
\begin{equation}
S(\vec{k}_{\mu},t)=
\frac{1}{\Delta x^d \,N^d}\langle|\hat{\phi}(\vec{k}_{\mu},t)|^2\rangle\,,
\label{eq:strkA}
\end{equation}
in terms of the Fourier transform of the field:
\begin{equation}
\hat{\phi}(\vec{k}_{\mu},t)=
(\Delta x)^{d}\sum_{j}e^{-i\vec{r}_j\cdot \vec{k}_{\mu}}\phi(\vec{r}_j,t)\,,
\end{equation}
and the correlation function $G(\vec r_i,t)$ is then computed as the inverse
Fourier transform of $S(\vec{k}_{\mu},t)$. 

We perform a spherical average of the pair correlation and 
structure functions
\begin{equation}
\label{sphericalg}
G(r,t)= \frac{1}{N_r}\sum_{r \le r_i < r+\Delta r}G(\vec{r}_i,t)
\end{equation}
\begin{equation}
\label{sphericals}
S(k,t)= \frac{1}{N_{k}}\sum_{k \le k_{\mu} < k+\Delta k}S(\vec{k}_{\mu},t)\,,
\end{equation}
where the sums run over the set of lattice points ($N_r$ and $N_k$)
between two circles of radius
$r$ and $r+\Delta r$, or $k$ and $k+\Delta k$, in real and reciprocal space,
respectively.

Finally, in order to compare results from different parameter values,
we have used the following normalization
\begin{equation}
g(r,t)= \frac{G(r,t)}{G(0,t)}\,,\qquad\qquad
s(k,t)= \frac{S(k,t)}{G(0,t)}\,\,.
\label{normals}
\end{equation}

When the interface between domains is very thin compared with their size,
the system has a single characteristic length
$R(t)$, which is related to the average size of domains. The scaling
hypothesis states that the pair correlation function depends on time only
through the time-dependent characteristic length:
\begin{equation}
g(r,t)=g(r/R(t))
\label{scalhyp}
\end{equation}
The length $R(t)$ can be defined in several ways, but in the scaling regime
all of them should lead to the same law for domain
growth. We have chosen $R(t)$ as the distance at which the pair correlation 
function is half its maximum value, i.e. $g(R(t),t)=1/2$. Since the structure
factor is the Fourier transform of the correlation function, we derive its
scaling law as
\begin{equation}
s(k,t)=R(t)^d\,s(k R(t))
\end{equation}
with no other explicit dependence on time.


\begin{figure}[htb]
\begin{center}
\epsfig{file=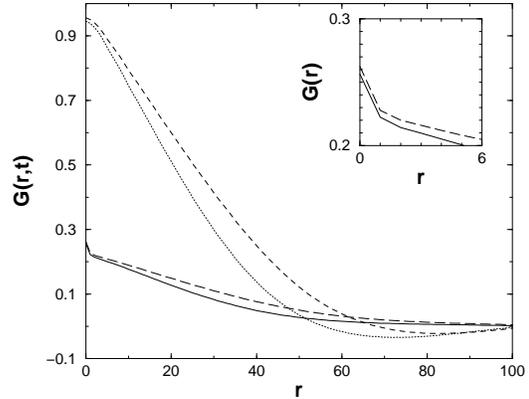, width=2.7in}
\end{center}
\caption{
Pair correlation function for the 2-dimensional non-conserved
Ginzburg--Landau model in the deterministic and stochastic cases.
The dotted (t=300) and dashed (t=500) lines correspond to the
deterministic case with $a=1$ and $D=1$.
The solid (t=300) and the long-dashed (t=500) lines correspond to the
stochastic model with $\epsilon=10^{-4}$, $\sigma^2=0.6$, $a=-0.2$
and $D=1$. In presence of noise sources
the pair correlation function exhibits a peak at $r=0$, as shown more
clearly in the inset.}
\label{fig:grtA}
\end{figure}



\begin{figure}[htb]
\begin{center}
\epsfig{file=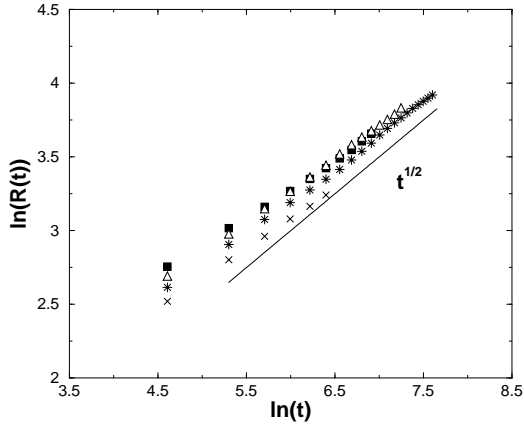, width=2.7in}
\end{center}
\caption{The Allen--Cahn law. Stars correspond to the deterministic
model with $a=1$ and $d=1$. Crosses correspond to the additive case
($\epsilon=0.7$, $a=2$ and $D=1$), squares to $\sigma^2=0.4$, and
triangles to $\sigma^2=0.6$. The two cases with 
multiplicative noise have $\epsilon=10^{-4}$, $a=-0.2$ and $D=1$.
The solid line is a guide to the eye.}
\label{fig:rtA}
\end{figure} 
 
We plot in Fig. \ref{fig:grtA} the pair correlation function for the
deterministic case with $a=1$ and for the nonequilibrium stochastic case with 
$\sigma^2=0.6$, $a=-0.2$, for two different times. As time increases, the correlation
fuction tends to a monotonically decreasing function with an increasing
characteristic decay length. A closer look (see inset of Fig. \ref{fig:grtA})
shows that in the stochastic case
the pair correlation function is not smooth near the origin.
This behaviour at short distances appears
whenever noise sources (either multiplicative or additive) exist,
and it differs from what is observed in the deterministic case, where the
pair correlation function is smooth at all distances.
As we will see, this behaviour is a bulk feature 
and, since it does not present scaling, we have avoided it in the following
section by substituting $G(r,t)$ near $r=0$ by a parabolic fit. 
The last section is devoted to the understanding of the behaviour of the
pair correlation function at short distances in the presence
of noise sources.


\begin{figure}[htb]
\begin{center}
\epsfig{file=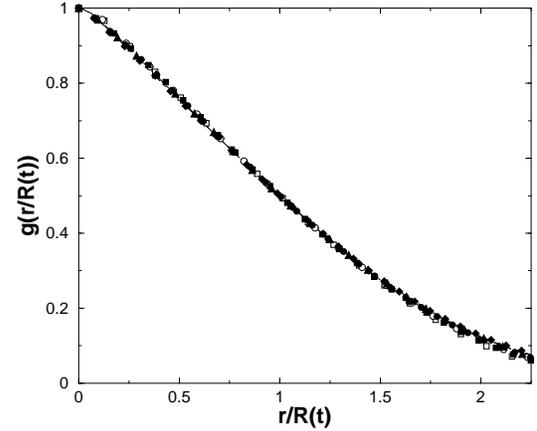, width=2.7in}
\vskip2mm
\centerline{(a)}
\vskip3mm
\epsfig{file=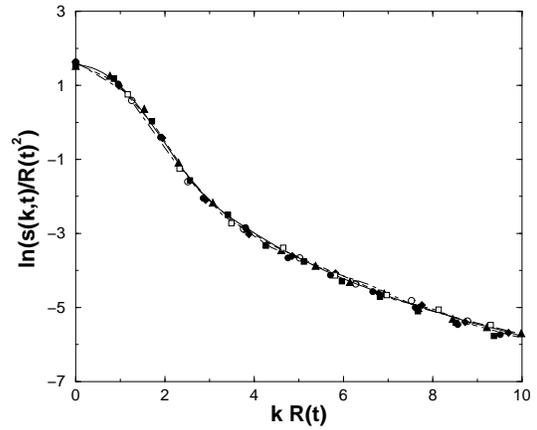, width=2.7in}
\vskip2mm
\centerline{(b)}
\end{center}
\caption{ 
Scaled pair correlation function (a) and structure funtion (b)
for the non-conserved Ginzburg--Landau model. Broken lines correspond to the 
deterministic model with $a=1$ and $D=1$ for t=1000 (dashed line) and 
t=1500 (dot--dashed line). Empty symbols, squares (t=500) and circles (t=600), 
correspond to the 
additive model  ($\epsilon=0.7$, $a=2$ and $D=1$). Full symbols correspond to 
the nonequilibrium models. Squares (t=800) and circles (t=1000) correspond
to $\sigma^2=0.4$, while triangles (t=600) and diamonds (t=1000) correspond
to $\sigma^2=0.6$.
Both cases have $a=-0.2$, $D=1$ and $\epsilon=10^{-4}$.
The continuous line in (b) is the theoretical prediction of
Ohta et al. \cite{Ohta82}} 
\label{fig:scalA}
\end{figure} 

Figure \ref{fig:rtA} shows the square root dependence on time of the
average domain size. The Allen-Cahn law is seen to be verified by all
models under study. The prefactor seems to depend on the noise intensity.

Numerical results for the scaled pair correlation and the structure
functions are shown in Figs. \ref{fig:scalA}(a) and \ref{fig:scalA}(b),
respectively. Our numerical simulations verify scaling for both the pair
correlation and the structure functions. Furthermore, the observed scaled
structure functions agree with the theoretical prediction of Ref.
\cite{Ohta82}, which has two adjustable parameters (a scale factor in
each axis), and are also in agreement with Ref. \cite{Oguz90}. 

\section{Conserved Model}

The conserved version of the model studied in the previous section, Eq.
(\ref{LangevinA}), is
\begin{equation}
\frac{\partial \phi(\vec x,t)}{\partial t} = -\nabla^2\left[a\phi-
\phi^3 +\,D\,
\nabla^2\phi+ \phi\,\xi(\vec x,t)\right] + \eta(\vec x,t)\,,
\label{LangevinB}
\end{equation}
which in the absence of external noise is the Cahn-Hilliard-Cook model
\cite{chc} and in the literature of critical phenomena \cite{hh} is known as
model B. The conservation law is $V^{-1}\int_V d\vec x \phi(\vec x,t)=
\phi_0$, constant, where $V$ is the total volume of the
system. This constant is given by the initial volume fraction of the system.
This conserved model is suitable for the description of the evolution of
systems such as binary alloys, in which the total concentration of each
component of the alloy is kept constant. In this case, the local order
parameter, $\phi(\vec x,t)$ represents the local difference of 
concentrations of each component of the alloy.  
As before, we consider external fluctuations on the control parameter $a$. 
Both additive and multiplicative noises are Gaussian, with zero mean
and correlation given by
\begin{equation}
\langle \eta(\vec x,t)\,\eta(\vec x,t')\rangle=-2\,\varepsilon\,\nabla^2\,
\delta(\vec x-\vec x')\,\delta(t-t')\,.
\label{aditivoB}
\end{equation}
and Eq.~(\ref{multiplicativoA}). The Laplacian term in Eq.~(\ref{aditivoB})
ensures that in absence of multiplicative noise and if $\epsilon=k_B T$
(the fluctuation--dissipation theorem), the stationary distribution is
still governed by a (restricted) Boltzmann--Gibbs distribution $\exp[-{\cal
F}/k_B\,T]\delta(V^{-1}\int_V d\vec x \phi(\vec x,t)-\phi_0)$. Let us consider
this equilibrium conserved model in a high--temperature, disordered, one--phase
state corresponding to a sufficiently low value of the parameter $a$. If we now suddenly decrease the temperature of the system
below its transition value (equivalently, by increasing $a$), the disordered one--phase state becomes unstable
and the system develops domains of the new phases which  slowly tend to the
equilibrium ones. The classical theory of the kinetics of first order phase
transitions distinguishes two regimes for the initial instability leading to
coarsening: for a quench taking the system
deep inside the coexistence curve (critical quench or
small order parameter $\phi_0$) the system is unstable against long-wavelength
perturbations and undergoes a process of
spinodal decomposition. On the other hand, for
quenches close to the coexistence curve, the system evolves by nucleation and
growth of the nuclei formed. In any case, since the order parameter is
conserved, the growth of a given domain is at expense of the smaller
domains of the same phase.
Domain growth appears as a result of diffusion accross the
interface between domains, caused by the interface curvature.
The equilibrium final state is a two--phase coexisting state. 
In this time regime, the average domain size grows with time as
$R(t)\propto \,t^{1/3}$,
which is called the Lifshitz--Slyozov law. 
This power--law behaviour has been derived analytically for small 
volume fractions, although it has been seen numerically to be also satisfied
even for large volume fractions \cite{TCG89}.
As in the non-conserved case, when the growth of domains
verifies the corresponding power--law behaviour, the pair correlation
function and the structure function satisfy the scaling hypothesis defined
in Eq.~(\ref{scalhyp}). The scaled functions have not been analytically
obtained so far. 

\begin{figure}[htb]
\begin{center}
\epsfig{file=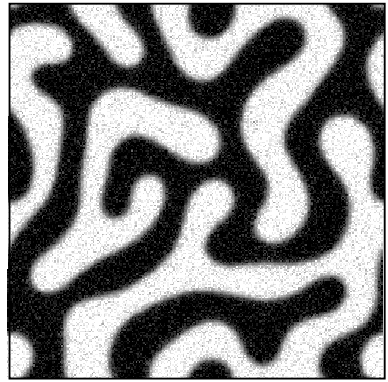, width=2.7in}
\end{center}
\caption{Spatial pattern for the stochastic model B at t=10000, with
multiplicative noise intensity $\sigma^2=0.1$,
$\epsilon=10^{-4}$, $a=-0.2$, $D=1$ and $\phi_0=0$. The square lattice has
$256\times 256$ cells of mesh size $\Delta x=1$. The black and white domains
correspond to symmetric phases.} 
\label{pat:stocB}
\end{figure}

In this section we are concerned with the dynamics towards the nonequilibrium,
two-phase coexisting steady state induced by external
fluctuations. Let us now consider the above nonequilibrium model [Eq.
(\ref{LangevinB})] in a disordered, $\phi_0=0$, one--phase steady state
corresponding to a small intensity of the multiplicative noise. 
If we now increase abruptly the intensity of the external fluctuations,
$\sigma$, the system develops domains of new phases that grow and tend to
the new stationary values (see Fig. \ref{pat:stocB}). For an external
noise with a white spatial correlation in the lattice ($c(0)=1$, as in
the previous section), a linear stability
analysis \cite{linstab} shows that
order appears in this case for values for $a$ above a critical
point $a_c=-2d\,\sigma^2$. Hence, for $-2d\,\sigma^2< a <0$, phase separation
induced by fluctuations appears. We will now address the issue
of whether this dynamics still verifies the Lifshitz--Slyozov law, and whether
a scaling behaviour for the pair correlation and structure functions exists.


As in the previous section, we define the model [given by Eq.
(\ref{LangevinB})] in a discrete space of mesh size $\Delta x$  
\begin{eqnarray}
\frac{d \phi_i}{ d t} = -\sum_k\,\widetilde D_{ik}\left[ a\,\phi_k - \phi_k^3+
\!D \sum_{j}
\,\widetilde D_{kj}\,\phi_j +\right.
\nonumber\\
\left.\phantom{\sum_{j}}
\phi_k\,\xi_k(t)\right]+ \eta_k(t)
\label{eq:spdedisB}
\end{eqnarray}
The discrete noises $\eta_i(t)$ and $\xi_i(t)$ are still Gaussian, with
zero mean and correlations given by
\begin{equation}
\langle \eta_i(t)\,\eta_j(t')\rangle=-2\,\varepsilon\,
\frac{\widetilde D_{i,j}}{\Delta x^d} \,\delta(t-t')
\label{eq:wnc2disB}
\end{equation}
and Eq.~(\ref{eq:wmn2dis}). These conserved noise term $\eta_i(t)$ can
be generated as the divergence of a stochastic vector field
\cite{noise_cons}.


\begin{figure}[htb]
\begin{center}
\epsfig{file=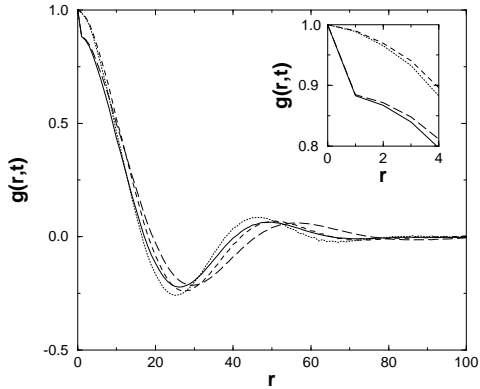, width=2.7in}
\end{center}
\caption{
Normalised pair correlation function for the 2--dimensional conserved model 
in the deterministic and nonequilibrium stochastic cases.
The dotted (t=7000) and dashed (t=10000) lines correspond to the
deterministic case with $a=2$ and $D=1$.
The solid (t=7000) and the long-dashed (t=10000) lines correspond to the 
stochastic model with $\epsilon=10^{-4}$, $\sigma^2=0.1$, $a=-0.2$ and $D=1$.
In presence of noise sources, the pair correlation function is not smooth
at short distances, as seen more clearly in the
inset. In both cases we have considered $\phi_0=0$.}
\label{fig:picoB}
\end{figure} 

We have studied two cases of noise-induced conserved
dynamics corresponding to two different intensities of the multiplicative
noise, $\sigma^2=0.1$ and $\sigma^2=0.2$. In both cases,
additive noise is $\epsilon=10^{-4}$ and the control parameter is
$a=-0.2$, which corresponds to a situation in which the disordeded
state, $\phi_i=0~\forall i$, is the deterministically stable phase. However,
since $-4 \sigma^2 < a < 0$, it turns out that for these multiplicative noise
intensities, the one--phase disordered state becomes unstable \cite{jordi98}.
We have compared their behaviour with three equilibrium cases:
the purely deterministic
model for $a=2$ and $a=0.2$, and the stochastic model B with only
additive noise ($\sigma=0$, $a=2$ and $\epsilon=0.7$). In all cases we
have taken the coupling coefficient as $D=1$.

We have performed numerical simulations of the full model in a two dimensional
regular square lattice of $256\times 256$ points with periodic boundary
conditions and mesh size $\Delta x=1$. Since the model is self--averaging
and the size we have chosen is relatively large, it has not been necessary
to consider many different realizations in order to have small statistical
errors. In particular, we have calculated averages over up to $6$ realizations
for the equilibrium models, and over $10$ realizations for the nonequilibrium
stochastic models 
with multiplicative noise. We have implemented a first--order Euler algorithm  
and used a time step of $dt=10^{-2}$ in the deterministic case and
$dt=5\cdot 10^{-3}$ in presence of noise sources. As initial conditions we
have considered a Gaussian random field with zero mean and variance
$10^{-2}$ ($\phi_0=0$).


\begin{figure}[htb]
\begin{center}
\epsfig{file=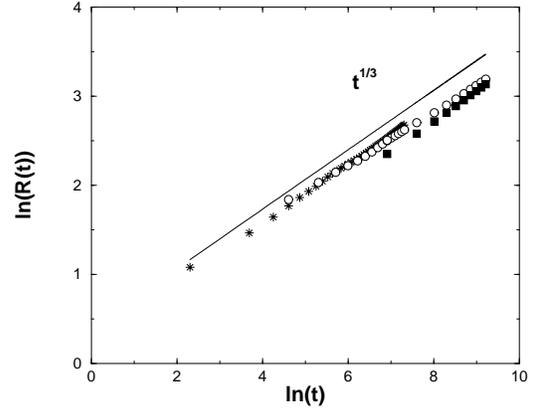, width=2.7in}
\end{center}
\caption{
Average domain size $R(t)$ for the conserved model.
Stars correspond to the deterministic case with $a=2$ and empty circles
to the stochastic equilibrium model with $a=2$ and $\epsilon=0.7$. Full
squares correspond to the nonequilibrium stochastic model with $\sigma^2=0.2$,
$\epsilon=10^{-4}$ and $a=-0.2$. The solid line is a guide to the eye. In all
cases the Lifshitz-Slyozov power law behaviour seems to be satisfied.}
\label{fig:snB}
\end{figure} 

The definition of the pair correlation function and the structure factor
are the same as in the non-conserved case. The only difference is that
now the spatial average of $\phi(\vec x,t)$ is a constant determined by
the initial condition. The pair--correlation function oscillates and decays
to zero with a characteristic distance that is a measure of the typical
domain size. This is shown in Fig. \ref{fig:picoB}, where we plot the
evolution of the pair correlation function for the deterministic ($a=0.2$) and 
stochastic ($\epsilon=10^{-4}$, $\sigma^2=0.1$, $a=-0.2$) cases.
The first zero crossing defines
the growth length, $g(R(t),t)=0$. As in the non-conserved case, we have
also found the same behaviour of
the correlation function at short distances that differs from what is 
observed in the deterministic case (see inset of Fig. \ref{fig:picoB}).
We have proceeded in a similar way as 
before by using a parabolic fit of $G(r,t)$ at $r=0$.

\begin{figure}[htb]
\begin{center}
\epsfig{file=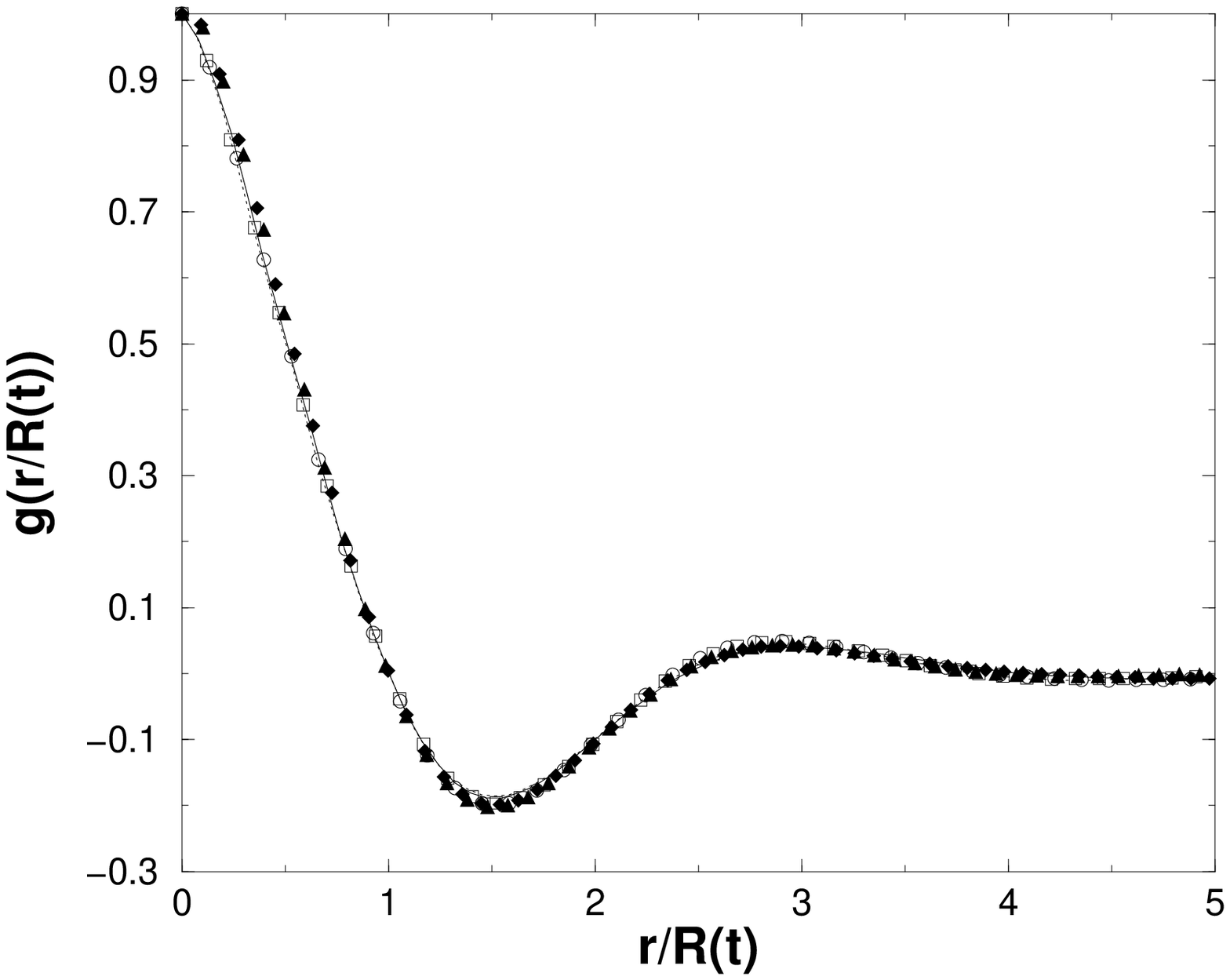, width=2.7in}
\centerline{(a)}
\epsfig{file=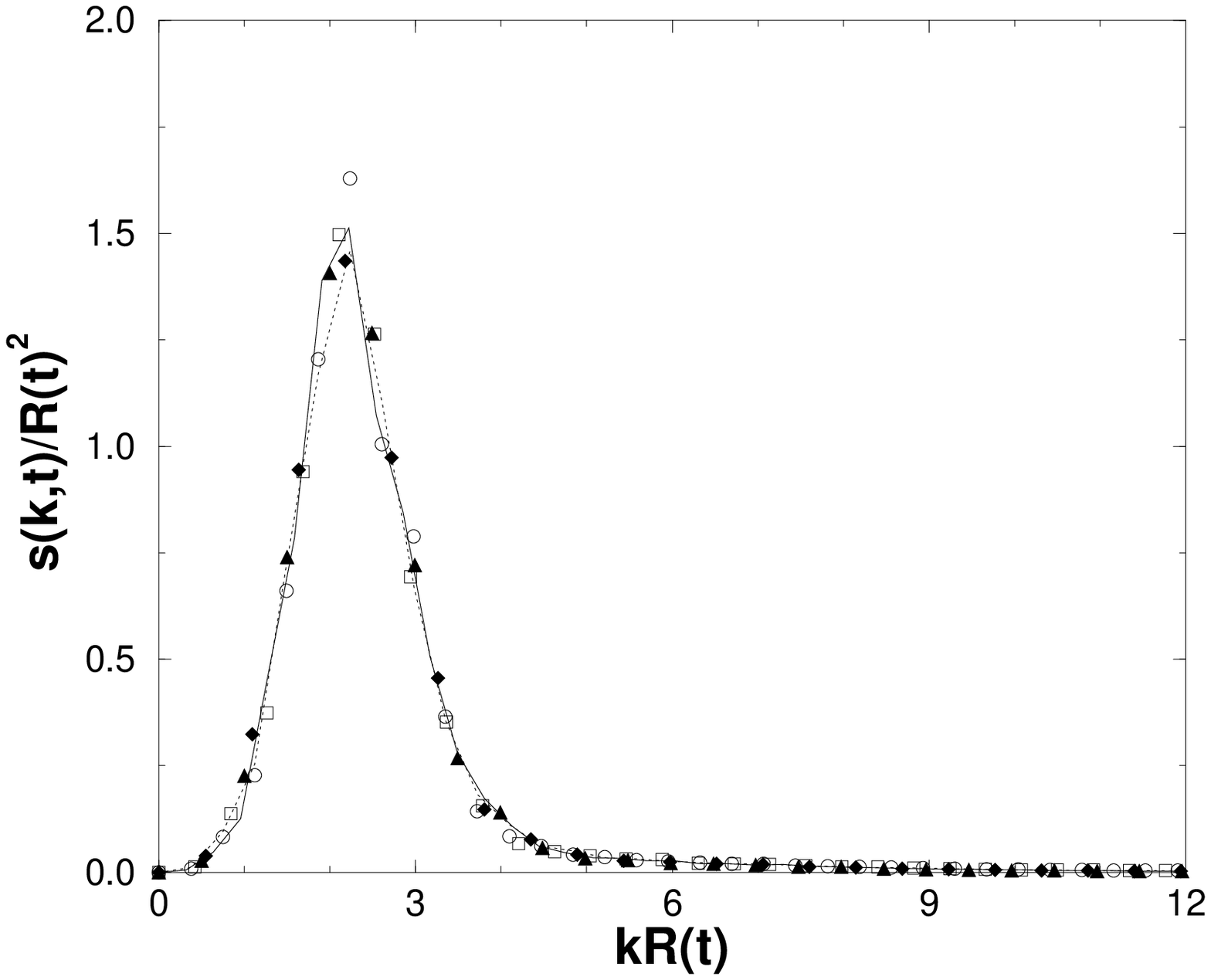, width=2.7in}
\centerline{(b)}
\end{center}
\caption{
Scaled pair correlation function (a) and scaled structure function (b)
for the conserved model.  
Lines correspond to the deterministic model with $a=2$ for $t=1000$ 
(continuous line) and $t=1600$ (dashed line). Empty symbols correspond to 
the equilibrium stochastic model with additive noise ($a=2$ and $\epsilon=0.7$) 
for $t=2000$ (circles) and $t=3000$ (squares). Full symbols correspond to 
the nonequilibrium stochastic model with $\sigma^2=0.2$, $\epsilon=10^{-4}$ 
and $a=-0.2$ for $t=7000$ (triangles) and $t=9000$ (diamonds).
In all cases $D=1$ and $\phi_0=0$.}
\label{fig:scalB}
\end{figure} 

As shown in Fig. \ref{fig:snB}, our numerical results indicate that the
average domain size obeys the equilibrium Lifshitz--Slyozov law even in the
presence of multiplicative noise. The corresponding scaled pair correlation 
and structure functions are plotted in Figs. \ref{fig:scalB}(a) and
\ref{fig:scalB}(b). The scaling behaviour is seen to be verified by the
nonequilibrium model with $\sigma^2=0.2$. Furthermore, the scaled functions
agree with the equilibrium ones.

\section{Noise Effects on the Short-Distance Behaviour of the
Correlation Function}

As we have seen in the previous sections, the discrete
two-point correlation
function of systems undergoing domain growth
in the presence of fluctuations is not smooth at short distances,
for both conserved and non-conserved dynamics 
(Figs. \ref{fig:grtA} and \ref{fig:picoB}).
This generic feature can be found in the existing literature \cite{Chak92}, 
but as far as we know, it has not been explained yet.
As we will see in this section, this behaviour is mainly a bulk feature, 
and the interfaces do not play any role. Therefore, it can be
studied considering the bulk stationary situation.

\subsection{Non-conserved Model}

\begin{figure}[htb]
\begin{center}
\epsfig{file=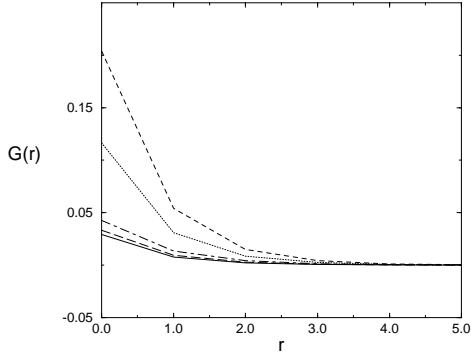, width=2.5in}
\centerline{(a)}
\epsfig{file=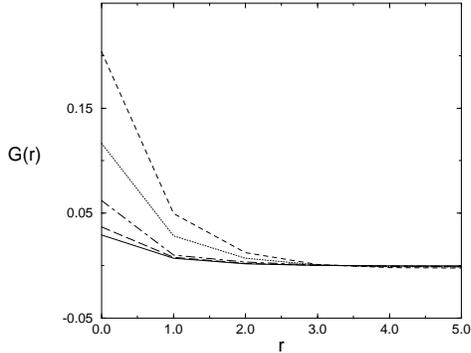, width=2.5in}
\end{center}
\centerline{(b)}
\caption{
Stationary pair correlation function for the non-conserved (a) and the
conserved  (b)
linear models for several noise intensities: only additive noise
with $\epsilon=0.1$ (solid line),  
$\epsilon=0.4$ (dotted line) and
$\epsilon=0.7$ (dashed line), additive ($\epsilon=0.1$) and multiplicative
noises with  
$\sigma^2=0.2$ (long dashed line) and 
$\sigma^2=0.5$ (dotted-dashed line). In all cases $a=-2$ and $D=1$ and
for the conserved case $\phi_0=0$.}
\label{fig:picolineal}
\end{figure}

We intend to show that the presence of fluctuations in the steady bulk
solution is responsible for the non-smooth behaviour of the discrete
correlation
function. To that end, we linearize the model given by Eq.~(\ref{LangevinA})
around the bulk phase, which, in order to simplify the analysis, we take as the 
homogeneous null phase ($\phi=0$). Let us start with the non-conserved model
with only additive noise, and linearize it around $\phi=0$  
\begin{equation}
\frac{\partial \phi(\vec x,t)}{\partial t} = a\phi+\,D\,
\nabla^2\phi+ \eta(\vec x,t)\,,
\label{LpicoA}
\end{equation}
where the additive noise has zero mean and correlations given by Eq.
(\ref{aditivoA}). The stability of this model requires $a$ to be negative. 
We now look for the dynamical equation of the pair--correlation function 
$G(\vec r,t)$, defined as the continuum version of (\ref{eq:corrA})
($\vec r=\vec x-\vec x'$) \cite{gsms}:
\begin{equation}
\label{dgdtcont}
\frac{\partial G(\vec r,t)}{\partial t}=2a  \,G(\vec r,t) +
2\,D\,\nabla^2\,G(\vec r,t) +2\epsilon\,\delta(\vec r)\,,
\end{equation}
which has a delta contribution due to additive noise.
In the stationary state, this equation can be solved in a
straightforward way in Fourier space. The resulting integral is
divergent in $d=2$, indicating the absence of a continuum limit in this
case. On the other hand, the integral converges in $d=1$.
Numerical simulations of Eq. (\ref{LpicoA}) [see Fig. \ref{fig:picolineal}(a)]
show a strong dependence on the noise intensity of the slope of the
correlation function at $r=0$. Therefore, this is a non-scaling feature,
which can be characterized by studying the spatial derivative of $G(r,t)$
near the origin. From (\ref{dgdtcont}), we obtain
\begin{equation}
\label{solcont}
\left.\frac{d}{d r} G(r,t)\right|_{r=0^+}=-\frac{\epsilon}{2D}\,,
\end{equation}
where we have taken into account that the pair correlation function is 
symmetric, $G(\vec r,t)=G(-\vec r,t)$. This expression shows clearly
the relevance of the noise intensity $\epsilon$. 

After this back-of-the-envelope calculation,
we perform a more exhaustive analysis in discrete space, in order to
compare it with the numerical results shown in Fig. \ref{fig:picolineal}(a). 
We now consider both additive and multiplicative noise sources and take the
bulk phase $\phi=0$. Thus, we linearize Eq.~(\ref{eq:spdedis}) around this
phase and obtain
\begin{equation}
\frac{d \phi_i}{ d t} = a\,\phi_i + \!D \sum_{j}
\,\widetilde D_{ij}\,\phi_j \, + \phi_i\,\xi_i(t)\, + \eta_i(t)\,.
\label{eq:1dAdis}
\end{equation}
From this equation, and making use of Novikov's theorem \cite{nises},
we can obtain the dynamical equation for the discrete
correlation $\langle \phi_i(t)\phi_j(t)\rangle$:
\begin{eqnarray}
\frac{d}{dt}\langle\phi_i (t)\phi_j(t)\rangle=2a\langle\phi_i\phi_j\rangle+
D \sum_{k}\,\left(\widetilde D_{ik}
\langle\phi_j\phi_k\rangle \,+\right. 
\nonumber\\
\left.
\widetilde D_{jk}\langle\phi_i\phi_k\rangle\right) +\,2\sigma^2\,
\langle\phi_i\phi_j\rangle\frac{\delta_{ii}+\delta_{ij}}{\Delta x^d}
+2\epsilon\frac{\delta_{ij}}{\Delta x^d} \,. 
\label{phiiphijev}
\end{eqnarray}
Since we are analyzing a bulk feature, the dynamics is no longer relevant,
and we can perform our calculations in steady state. Therefore,
we set $\frac{d}{dt}\langle\phi_i (t)\phi_j(t)\rangle=0$,
for any $i$ and $j$. Since we are interested on very short distances,
we focus on $i=j$, to obtain:
\begin{eqnarray}
a\langle\phi_i^2\rangle+\,\frac{D}{\Delta x^2} \sum_{nn(i)}\left(
\langle\phi_i\phi_{nn(i)}\rangle-\langle\phi_i^2\rangle\right)\,+
\nonumber\\
2\frac{\sigma^2}{\Delta x^d}\langle\phi_i^2\rangle+
\frac{\epsilon}{\Delta x^d}=0\,,
\label{phiphid}
\end{eqnarray}
where the Laplacian term has been made explicit and the sum runs over the 
$2d$ nearest neighbors of $i$. 

In discrete space, the first derivative of the two-point correlation function 
$G_i\equiv G(\vec r_i)$ at $i=0$ is
\begin{equation}
G'_0\equiv \frac{\langle\phi_i\phi_{nn(i)}\rangle-\langle\phi_i^2\rangle}
{\Delta x}\,.
\label{derivada}
\end{equation}
Taking into account the isotropy of the system, we have
\begin{equation}
\langle\phi_i\phi_{nn(i)}\rangle=G_1\,,
\hskip1cm
\forall \,\,\,nn(i)
\label{g1}
\end{equation}
where $G_1$ is the two-point correlation 
function for distances equal to one cell. 

Considering Eqs.
(\ref{derivada}) and
(\ref{g1}), expression (\ref{phiphid}) can be rewritten in the following form
\begin{equation}
G'_0= -\frac{a}{2d\,D}\Delta x\,G_0-
\frac{\sigma^2}{d\,D\Delta x^{d-1}}\,G_0-\frac{\epsilon}{2d\,D\Delta x^{d-1}}.
\label{discpicoA}
\end{equation}
If we now take the continuous limit $\Delta x\rightarrow 0$ and consider that 
only additive noise is present ($\sigma^2=0$), we recover the
continuous result (\ref{solcont}) given above in the case of $d=1$.

The numerical simulations shown in Fig. \ref{fig:picolineal}(a) have been
performed on a
one-dimensional lattice of $N=16384$ points and mesh size $\Delta x=1$,
for $a=-2$ and $D=1$.
We have used a first--order Euler scheme with $dt=5\cdot10^{-3}$, and
computed the two-point correlation function in the stationary state.


Figure \ref{fig:discont} shows numerical and theoretical results for
$G'_0=(G_1-G_0)/\Delta x$ in two different cases, the linear model with
only additive noise and with both additive and multiplicative noises.
Empty symbols are simulation results: squares correspond to different
additive noise intensities (bottom axis) and $\sigma^2=0$, and triangles
correspond to different multiplicative noise intensities (top axis) and
fixed additive noise $\epsilon=0.1$. Lines are the theoretical prediction
(Eq.~(\ref{discpicoA})). We have to note that our theoretical
prediction uses the simulation result for $G_0$.

\subsection{Conserved model}

Following the same scheme as in the previous section, we now analyze
the same phenomena for the conserved model. By linearizing
Eq.~(\ref{LangevinB}) with only additive noise around $\phi(\vec r,t)=0$,
we obtain 
\begin{equation}
\frac{\partial \phi(\vec x,t)}{\partial t} = 
-\nabla^2\left[a\phi+\,D\,\nabla^2\phi \right]
+ \eta(\vec x,t)\,,
\label{LpicoB}
\end{equation}
where $a$ must be negative and the additive noise is given by Eq.
(\ref{aditivoB}). The corresponding dynamical equation for the
pair correlation function is
\begin{eqnarray}
\label{dgdtcontB}
\frac{\partial G(\vec r,t)}{\partial t}=-\nabla^2\,\left[2a  \,G(\vec r,t) +
2\,D\,\nabla^2\,G(\vec r,t) +
\right.
\nonumber\\
\left.
2\epsilon\,\delta(\vec r)\right]\,.
\end{eqnarray}
In the stationary state we have
\begin{equation}
\label{gstcontB}
a  \,G(\vec r,t) +
\,D\,\nabla^2\,G(\vec r,t) +\epsilon\,\delta(\vec r)=h
\end{equation}
where h is a constant since the steady state is homogeneous. In the case
$\phi_0=0$, this constant is zero \cite{Ibanes99,Munoz98}.
Comparing Eq.~(\ref{gstcontB}) with Eq.~(\ref{dgdtcont}) in the stationary case,
it can be seen that also in the conserved case Eq.~(\ref{solcont}) holds
in $d=1$.

In a more detailed analysis, we consider the lattice case when multiplicative
noise is also present. 
We linearize Eq.~(\ref{eq:spdedisB}) around $\phi=0$ to obtain: 
\begin{eqnarray}
\frac{d \phi_i}{ d t} = -\sum_k\widetilde D_{ik}\!\left(a\,\phi_k + \!D \sum_{j}
\,\widetilde D_{kj}\,\phi_j  + \phi_i\,\xi_k(t)\right)
\nonumber
\\
+ \eta_i(t),\;\;
\label{eq:spdispico}
\end{eqnarray}
where the discrete noises $\eta_i(t)$ and $\xi_i(t)$ are 
Gaussian with zero mean and correlation given by Eqs.(\ref{eq:wnc2disB}) and 
(\ref{eq:wmn2dis}).

The procedure of the analysis is the same as in the non-conserved case.
First we look for the dynamical equation of the two-point correlation
function, which using Novikov's theorem \cite{nises}, can be derived to be: 
\begin{eqnarray}
\label{2momentB}
\frac{d\,}{dt}\langle \phi_i\phi_j\rangle=
\sum_s\widetilde D_{is}\!\left[-a\langle \phi_j\phi_s\rangle
\!-\!D\sum_k\widetilde D_{sk}\langle \phi_k\phi_j\rangle+ \right.
\nonumber\\
\left.
\,\sigma^2\sum_k \frac{\delta_{sk}}{\Delta x^d}
\left(\widetilde D_{sk}\langle\phi_k\phi_j\rangle+
\widetilde D_{jk}\langle\phi_k\phi_s\rangle\right)\right]\,+
\nonumber\\
\sum_s\widetilde D_{js}\left[-a\langle \phi_i\phi_s\rangle 
-D\sum_k\widetilde D_{sk}\langle \phi_k\phi_i\rangle\,+ \right. 
\nonumber\\
\left.
\,\sigma^2\sum_k \frac{\delta_{sk}}{\Delta x^d}
\left(\widetilde D_{sk}\langle\phi_k\phi_i\rangle+
\widetilde D_{ik}\langle\phi_k\phi_s\rangle\right)\right]
-2\frac{\epsilon\widetilde D_{ij}}{\Delta x^d}.\;\;
\end{eqnarray}
We now focus on the stationary state $\frac{d\,}{dt}\langle 
\phi_i\phi_j\rangle=0$ and consider the case $i=j$
\begin{eqnarray}
\label{2momentiistB}
2\sum_s\widetilde D_{is}\left[a\langle \phi_i\phi_s\rangle
+D\sum_k\widetilde D_{sk}\langle \phi_k\phi_i\rangle
+\epsilon\frac{\delta_{is}}{\Delta x^d}
\right. 
\nonumber\\
\left.
\,-\sigma^2\sum_k \frac{\delta_{sk}}{\Delta x^d}
\left(\widetilde D_{sk}\langle\phi_k\phi_i\rangle+
\widetilde D_{ik}\langle\phi_k\phi_s\rangle\right)
\right]=0\,,
\end{eqnarray}
where the additive noise term has been introduced inside the Laplacian.
This equation is satisfied if each term inside the Laplacian is equal to a 
constant $h$.
\begin{eqnarray}
\label{h}
a\langle \phi_i\phi_s\rangle
+D\sum_k\widetilde D_{sk}\langle \phi_k\phi_i\rangle
+\epsilon\frac{\delta_{is}}{\Delta x^d}
\nonumber\\
\,-\sigma^2\sum_k \frac{\delta_{sk}}{\Delta x^d}
\left(\widetilde D_{sk}\langle\phi_k\phi_i\rangle+
\widetilde D_{ik}\langle\phi_k\phi_s\rangle\right)=h
\end{eqnarray}
for any $s,i$. As before, for $\phi_0=0$, the constant $h$ must be zero.  
We now take the case $s=i$. 
If we use definition (\ref{derivada}) for the discrete derivative of the 
pair correlation function and take into account the 
isotropy of the system, Eq.~(\ref{g1}), we can obtain from Eq.~(\ref{h})
\begin{equation}
G'_0= -\frac{a\, G_0 \Delta x}{2d\,D}-
\frac{2\sigma^2\,G_0}{D\Delta x^{d+1}}-\frac{\epsilon}{2d\,D\Delta x^{d-1}}
-\frac{h\Delta x}{2d\,D}\,,
\label{discpicoB}
\end{equation}
which in the continuous limit $\Delta x\rightarrow 0$ and in absence of
external fluctuations ($\sigma^2=0$) coincides with the continuous analysis
of Eq. (\ref{solcont}) for $d=1$. Notice that the dependence on 
the additive noise intensity is the same as in the non-conserved model, 
while the effect of multiplicative noise is different. Both results can
be made to coincide if the intensity of the multiplicative noise of the
non-conserved model A, $\sigma^2_A$ is related to the one of the conserved
model B, $\sigma^2_B$, by the relation $\sigma^2_A=(2d/\Delta x^2)\sigma^2_B$.
This relation between both models in presence of multiplicative noise has
been established in several analyses \cite{jordi98,Ibanes99}. 


\begin{figure}[htb]
\begin{center}
\epsfig{file=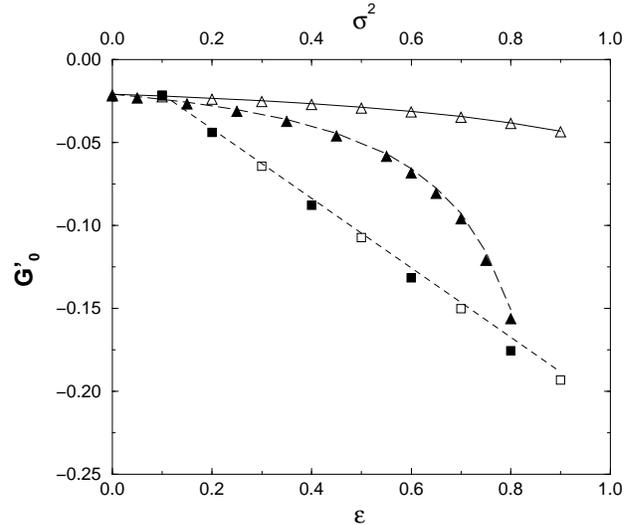, width=3.2in}
\end{center}
\caption{Discrete first derivative $G'_r$ of the pair correlation
function in the stationary state at $r=0$ for both non-conserved 
and conserved models with $a=-2$ and $D=1$. 
Symbols are simulation results for both the non-conserved (empty symbols)
and the conserved models (full symbols) 
and lines are the theoretical predictions.  
Squares correspond to different additive noise intensities $\epsilon$
(bottom axis) with $\sigma^2=0$ and
triangles to different multiplicative noise intensities $\sigma^2$ (top axis) 
with fixed additive noise 
$\epsilon=0.1$.
For the conserved model we have taken $\phi_0=0$.}
\label{fig:discont}
\end{figure} 

Again, we have performed simulations of Eq.~(\ref{eq:spdispico}) with
$a=-2$ and $D=1$ for $\phi_0=0$ with only additive noise,
and with both white additive and multiplicative noises in a one dimensional 
lattice of mesh size $\Delta x=1$ and $N=16384$ cells. We have used a 
first--order Euler algorithm with time step $dt=5\cdot 10^{-3}$.  
Results for different cases of the stationary correlation function are
shown in Fig. \ref{fig:picolineal}(b). When only additive noise
is present, the pair correlation function of both conserved and 
non-conserved models coincide [cf. Fig. \ref{fig:picolineal}(a)].
As in the non-conserved case, the contribution of additive noise is here
more important than the corresponding to multiplicative noise.

Analytical and numerical results for $G'_0$ are shown in Fig.
\ref{fig:discont}. Full symbols are the simulation results in this case:
squares correspond again to different additive noise intensities
(bottom axis) with $\sigma^2=0$, and triangles to
different multiplicative noise intensities (top axis) with fixed additive
noise $\epsilon=0.1$. Lines are the corresponding theoretical predictions
[Eq.~(\ref{discpicoB}) with $h=0$] using the values of $G_0$ obtained from
the simulations.

\section{Conclusions}

In this paper we have studied the dynamics of phase ordering that appears
in a non-potential situation induced by external noise. We have
considered both non-conserved (order-disorder) and conserved
(phase-separation) situations. Concerning the time
evolution of the characteristic size $R(t)$ of the domains, our results
indicate that $R(t)$ grows with time as a power law. The exponent of the
power law is $1/2$ for the non-conserved case and $1/3$ for the conserved
case. These two values coincide with the equivalent ones obtained in the
decay towards an {\em equilibrium} state (either deterministic or in the
presence of internal fluctuations). We have found a scaling description
consistent with the fact that $R(t)$ is the only relevant length scale of
the system, similarly also to the equilibrium cases.
Moreover, the corresponding scaling functions are in agreement with
the ones observed in the decay towards equilibrium.
These results indicate that the physical mechanisms underlying noise-induced
phase-ordering processes are identical to the deterministic ones, namely
interface curvature in the non-conserved case, with the addition of diffusion
in the conserved situation \cite{pelce}. Finally, we
have analyzed in detail a non-smooth, non-scaling behaviour induced by
fluctuations that appears
in the pair correlation function at short distances. We have shown that
this behaviour is due to fluctuations in the bulk, and provided a
theoretical expression for its magnitude. 

\section{Acknowledgements}

This work has been supported by the Direcci\'on General de
de Ense\~nanza Superior e Investigaci\'on Cient\'{\i}fica
(Spain), under projects PB94--1167, PB96--0241,
PB97--0141-C02-01, and PB98-0935. M.I. also acknowledges the Direcci\'on General
de Ense\~nanza Superior e Investigaci\'on Cient\'{\i}fica for
financial support.


\end{document}